\documentclass[10pt, conference]{IEEEtran}

\IEEEoverridecommandlockouts
\usepackage{amsthm}
\usepackage{amsmath,amssymb}
\usepackage[caption=false,font=normalsize,labelfont=sf,textfont=sf]{subfig}
\usepackage{thmtools}
\usepackage{xcolor}

\usepackage{bm}
\allowdisplaybreaks

\usepackage{algorithmic}
\usepackage{graphicx}
\usepackage{textcomp}
\usepackage[draft=true]{hyperref}
\usepackage{cleveref}
\usepackage{commath}
\usepackage{optidef}
\usepackage{orcidlink}
\usepackage{stfloats}
\usepackage{cite}
\usepackage{enumitem}

\hypersetup{
  bookmarks=false,
  colorlinks=true,
  linkcolor=black,
  citecolor=black,
  urlcolor=black,
  pdfborder={0 0 0},
  breaklinks=true
}

\usepackage[left=1.62cm,right=1.62cm,top=1.7cm,bottom=2.67cm]{geometry}
\IEEEaftertitletext{\vspace{-0.1\baselineskip}}
\creflabelformat{lemma}{#2\textbf{#1}#3}
\creflabelformat{theorem}{#2\textbf{#1}#3}
\creflabelformat{corollary}{#2\textbf{#1}#3}
\creflabelformat{proposition}{#2\textbf{#1}#3}
\crefname{lemma}{\textbf{Lemma}}{\textbf{Lemmas}}
\crefname{theorem}{\textbf{Theorem}}{\textbf{Theorems}}
\crefname{corollary}{\textbf{Corollary}}{\textbf{Corollaries}}
\crefname{proposition}{\textbf{Proposition}}{\textbf{Propositions}}
\crefname{section}{Section}{Sections}
\crefname{figure}{Fig.}{Figs.}

\title{On Energy-Efficient Passive Beamforming Design\\of RIS-Assisted CoMP-NOMA Networks}

\author{
    \IEEEauthorblockN{
        Muhammad Umer\IEEEauthorrefmark{1},
        Muhammad Ahmed Mohsin\IEEEauthorrefmark{1},
        Aamir Mahmood\IEEEauthorrefmark{2},
        Haejoon Jung\IEEEauthorrefmark{3},
        Haris Pervaiz\IEEEauthorrefmark{4},\\
        Mikael Gidlund\IEEEauthorrefmark{2},
        and Syed Ali Hassan\IEEEauthorrefmark{1}}
    \IEEEauthorblockA{\IEEEauthorrefmark{1}School of Electrical Engineering and Computer Science (SEECS),\\National University of Sciences and Technology (NUST), Pakistan}
    \IEEEauthorblockA{\IEEEauthorrefmark{2}Department of Computer and Electrical Engineering, Mid Sweden University, 851 70 Sundsvall, Sweden}
    \IEEEauthorblockA{\IEEEauthorrefmark{3}Department of Electronics and Information Convergence Engineering, Kyung Hee University, Republic of Korea}
    \IEEEauthorblockA{\IEEEauthorrefmark{4}School of Computer Science and Electronic Engineering (CSEE), University of Essex, UK}
    \IEEEauthorblockA{Email: \{mumer.bee20seecs, mmohsin.bee20seecs, ali.hassan\}@seecs.edu.pk, \\\{aamir.mahmood, mikael.gidlund\}@miun.se, haejoonjung@khu.ac.kr, haris.pervaiz@essex.ac.uk
    }
}

\begin{document}
\maketitle

\begin{abstract}
  This paper investigates the synergistic potential of reconfigurable intelligent surfaces (RIS) and non-orthogonal multiple access (NOMA) to enhance the energy efficiency and performance of next-generation wireless networks. We delve into the design of energy-efficient passive beamforming (PBF) strategies within RIS-assisted coordinated multi-point (CoMP)-NOMA networks. Two distinct RIS configurations, namely, enhancement-only PBF (EO) and enhancement \& cancellation PBF (EC), are proposed and analyzed. Our findings demonstrate that RIS-assisted CoMP-NOMA networks offer significant efficiency gains compared to traditional CoMP-NOMA systems. Furthermore, we formulate a PBF design problem to optimize the RIS phase shifts for maximizing energy efficiency. Our results reveal that the optimal PBF design is contingent upon several factors, including the number of cooperating base stations (BSs), the number of RIS elements deployed, and the RIS configuration. This study underscores the potential of RIS-assisted CoMP-NOMA networks as a promising solution for achieving superior energy efficiency and overall performance in future wireless networks.
\end{abstract}


\section{Introduction}
The escalating demand for ubiquitous connectivity and heightened data rates in future wireless networks compels the exploration of innovative solutions that transcend conventional methodologies. Reconfigurable intelligent surfaces (RIS) have emerged as a transformative technology, capable of dynamically manipulating electromagnetic wave propagation and customizing wireless environments to cater to specific communication requirements. By intelligently controlling the phase shifts of incident signals, RIS holds the potential to significantly enhance coverage, capacity, and spectral efficiency in a cost-effective and energy-efficient manner, making them a key enabler for Beyond 5G (B5G) and sixth-generation (6G) wireless networks~\cite{wu2021intelligent, zeng2020reconfigurable, zhang2020capacity, huang2019reconfigurable, you2020energy}.

However, the vision of future networks supporting massive connectivity with an unprecedented multitude of intelligent devices and significantly higher density (estimated at $10^6$ devices per km$^2$~\cite{akyildiz20206g}) presents challenges for traditional orthogonal multiple access (OMA) due to the limited availability of orthogonal resource blocks at each base station (BS). Non-orthogonal multiple access (NOMA) emerges as a promising solution, enabling multiple users to share the same time-frequency resources through superposition coding at the transmitter and successive interference cancellation (SIC) at the receiver~\cite{liu2017nonorthogonal}. This facilitates efficient utilization of the available spectrum and improved performance, particularly in terms of outage probability and spectral efficiency~\cite{ding2014performance, yue2018unified}.

Building upon the evolution of cellular networks, coordinated multi-point (CoMP) techniques have been standardized to address interference and spectrum limitations. The integration of CoMP and NOMA has gained significant interest due to its potential to further enhance spectral efficiency and user fairness in multi-cell environments. In CoMP-NOMA networks, multiple BSs act cooperatively to improve the performance of edge users, mitigate interference, and increase overall system capacity~\cite{ali2018downlink, elhattab2020comp}.
% However, coordination among all BSs presents challenges due to inaccurate channel state information (CSI), additional synchronization across cells, and increased signal processing requirements.

Recent research endeavors have explored the potential of RIS-assisted CoMP-NOMA networks, highlighting the synergistic advantages of these technologies. Strategic placement of RIS at the cell edge can significantly enhance signal quality for edge users, leading to improved coverage and capacity~\cite{elhattab2020reconfigurable}. Furthermore, RIS can be employed to simultaneously improve signal quality and suppress interference, thereby enhancing the performance of CoMP-NOMA networks~\cite{hou2021joint}. However, a crucial question remains largely unanswered: \textit{how do different RIS configurations, the number of cooperating BSs, and the number of RIS elements influence network energy efficiency?}

Motivated by this knowledge gap, this work delves into an energy-efficient passive beamforming (PBF) design of RIS-assisted CoMP-NOMA networks. We propose two distinct RIS configurations:

\begin{enumerate}[leftmargin=*]
  \item \textit{Enhancement-only PBF (EO):} Optimizes the RIS phases to enhance the desired signal quality for edge users.
  \item \textit{Enhancement \& Cancellation PBF (EC):} Optimizes the RIS phases for both signal enhancement and interference suppression.
\end{enumerate}

This work delves into a comprehensive analysis of network energy efficiency under diverse RIS configurations, exploring the influence of the number of cooperating BSs and RIS elements on overall system performance. We formulate a PBF design problem with the objective of maximizing energy efficiency through the optimization of RIS phase shifts. Our findings reveal that the integration of RIS with CoMP-NOMA networks yields substantial improvements in both energy efficiency and overall network performance, underscoring their potential for future wireless communication systems.

\section{System Model}

\subsection{System Description}
\begin{figure}
  \centering
  \includegraphics[width=0.925\columnwidth]{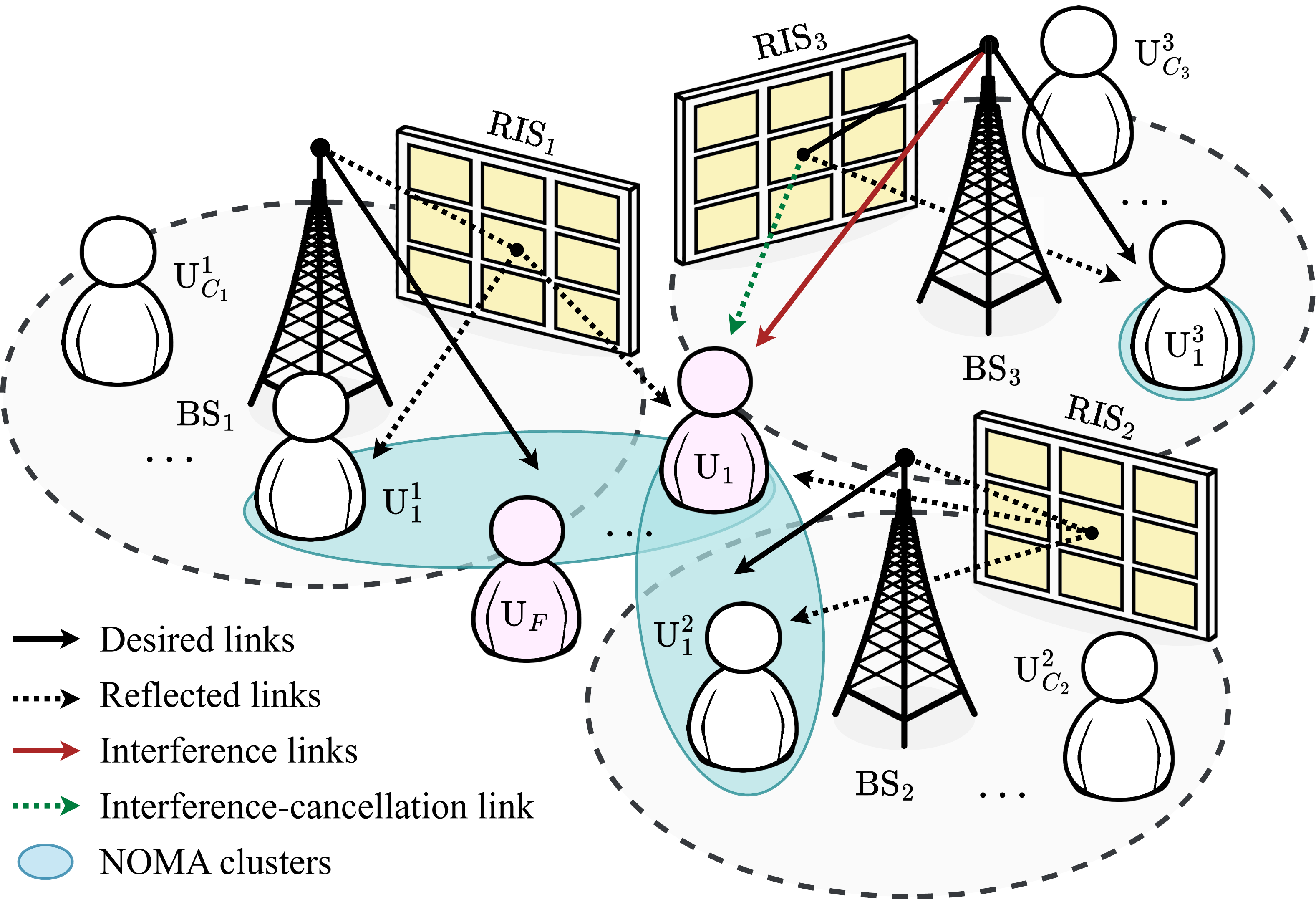}
  \caption{An illustration of the RIS-assisted multi-cell CoMP-NOMA network.}
  \label{fig:system}
  \vspace*{-0.25em}
\end{figure}

We consider a downlink transmission scenario in an RIS-assisted multi-cell CoMP-NOMA network, as depicted in Fig.~\ref{fig:system}. The network is comprised of $I$ cells, each modeled as a disk with radius $R_i$ and served by a BS located at its center, denoted by BS$_i$, where $i \in \mathcal{I} \triangleq \{1, 2, \ldots, I\}$. Each single-antenna BS utilizes two-user NOMA to serve user clusters, also equipped with single antennas, within its coverage area.\footnote{Due to processing complexity, latency of SIC at the receivers, and the practical limitations resulting in SIC error propagation, two-user NOMA pairs are considered in this work. Moreover, two-user configurations are of practical interest and have been standardized in 3GPP Release 15~\cite{ding2020simple}.}

% Change width to 0.9\columnwidth from 1\columnwidth
% to fix gutter error in EDAS

Users within the network are categorized into two distinct classes based on their spatial location: cell-center users and edge users. The cell-center users reside within the disk of their associated cell, while the edge users fall outside it. Let $\mathcal{C}^{(i)} \triangleq \{1, 2, \ldots, C_i\}$ represent the set of indices for cell-center users associated with BS$_i$, and $\mathcal{F} \triangleq \{1, 2, \ldots, F\}$ represent the set of indices for the shared edge users cooperatively served by multiple BSs.
% Here, $C_i$ and $F$ denote the cardinality (number of elements) of the respective sets.
We denote cell-center users as U$^i_c$ (superscript $i$ indicates the serving BS) and edge users as U$^e_f$ (superscript $e$ represents the edge class), with $c \in \mathcal{C}^{(i)}$ and $f \in \mathcal{F}$. Furthermore, $\mathcal{U} = \bigcup_{i \in \mathcal{I}} \mathcal{C}^{(i)} \cup \mathcal{F}$ represents the index set for all users in the network. For ease of exposition and without loss of generality, we consider a single edge user and a single cell-center user per cell in this work, i.e., $C_i = F = 1$, $\forall i \in \mathcal{I}$.

For coordinated operation, the BSs are assumed to be interconnected via a high-speed backhaul network to a central processing unit (CPU).
% This backhaul network facilitates information exchange between the BSs, enabling CoMP techniques for enhanced network performance.
BSs participating in CoMP are referred to as cooperative BSs and are denoted by the set $\mathcal{J} \triangleq \{1, 2, \ldots, J\}$, where $J \leq I$. To further improve the signal quality for edge users, each BS$_i$ is equipped with an RIS, denoted R$_i$, strategically placed at the cell edge.

\subsection{Channel Model}
Our analysis encompasses the impact of both large-scale and small-scale fading effects on each communication link within the system. The direct links, i.e., the channels between the BSs and the users, are characterized as Rayleigh fading channels due to the significant propagation distances and the presence of numerous scatterers in the environment. We denote the channel between BS$_i$ and user U$^n_u$ as $h^n_{i, u}$, where $n \in \{\mathcal{I}, e\}$ for cell-center or edge user, and $u \in \mathcal{U}$. Mathematically, this can be expressed as
\begin{equation}
  h^n_{i, u} = \sqrt{\frac{\rho_o}{PL(d^n_{i, u})}} v^n_{i, u},
\end{equation}
where $\rho_o$ is the reference path loss at $1$ m, $PL(d^n_{i, u})$ is the large-scale path loss modeled as $PL(d^n_{i, u}) = (d^n_{i, u})^{\alpha_n}$, such that $\alpha_n$ is the path loss exponent, $d^n_{i, u}$ is the distance between BS$_i$ and U$^n_u$, and $v^n_{i, u} \in \mathbb{C}^{1\times 1}$ is the small-scale Rayleigh fading coefficient with zero mean and unit variance.

In contrast to the Rayleigh fading experienced on direct links, the channels between BSs and RIS are modeled using Rician fading due to the presence of a dominant line-of-sight (LoS) component. The channel between BS$_i$ and R$_i$ is denoted $\mathbf{h}_{i, \text{R}_i}$ and can be expressed as
\begin{multline}
  \mathbf{h}_{i, \text{R}_i} = \\
  \sqrt{\frac{\rho_o}{PL(d_{i, \text{R}_i})}} \left( \sqrt{\frac{\kappa}{1 + \kappa}} \mathbf{g}^{\text{LoS}}_{i, \text{R}_i} + \sqrt{\frac{1}{1 + \kappa}} \mathbf{g}^{\text{NLoS}}_{i, \text{R}_i} \right),
\end{multline}
where $\kappa$ is the Rician factor, $\mathbf{g}^{\text{LoS}}_{i, \text{R}_i} \in \mathbb{C}^{K\times 1}$ is the LoS channel vector given by
\begin{equation*}
  \mathbf{g}^{\text{LoS}}_{i, \text{R}_i} = \left[1, \ldots, e^{j(k-1)\pi\sin(\omega_i)}, \ldots, e^{j(K-1)\pi\sin(\omega_i)}\right]^T,
\end{equation*}
where $k \in \{1, 2, \ldots, K\}$ indexes elements of R$_i$ and $\omega_i$ represents the angle of arrival (AoA) of the LoS component at R$_i$ while $\mathbf{g}^{\text{NLoS}}_{i, \text{R}_i} \in \mathbb{C}^{K\times 1}$ is the NLoS component which follows Rayleigh fading as previously described. The channel between RIS and edge users can be modeled similarly using Rician fading.

For simplicity, we assume perfect channel state information (CSI) at the BSs. While attaining perfect CSI in practical scenarios can be challenging, recent advancements in channel estimation techniques for RIS-assisted wireless networks have demonstrated the potential for accurate CSI acquisition with a reasonable overhead~\cite{zheng2022survey, wei2021channel, shtaiwi2021channel, zhou2022channel}.

\subsection{RIS Configuration}
Each BS$_i$ utilizes a passive RIS R$_i$ to improve signal quality or suppress interference for U$^e_f$, $\forall f \in \mathcal{F}$. We assume an optimal deployment location of the RIS at the cell edge, allowing for effective manipulation of signals for edge users. RIS elements can independently adjust the phase shift of the incident signal and are assumed to be controlled by the CPU. Furthermore, the phase shift (PS) matrix associated with R$_i$ is expressed as $\mathbf{\Theta}_i = \text{diag}(l_1 e^{j\theta^i_1}, l_2 e^{j\theta^i_2}, \ldots, l_K e^{j\theta^i_K})$, where $l_k \in (0, 1]$ is the amplitude adjustment factor and $\theta^i_k \in [-\pi, \pi)$ is the phase shift of the $k$-th element. In this work, we assume an ideal RIS with perfect phase control and all reflection elements having a unit amplitude ($l_k = 1, \forall k$).\footnote{While practical RIS implementations are subject to phase quantization errors, this work prioritizes establishing a proof-of-concept for the benefits of RIS-assisted CoMP-NOMA in terms of energy efficiency and performance.
  % Future work can explore incorporating phase quantization effects into the model for a more realistic analysis.
}

\section[Performance Analysis and Passive Beamforming Design]{Performance Analysis and \\Passive Beamforming Design}

\subsection{Rate and Outage Probability Analysis}
According to the NOMA principle, the BSs serve multiple users simultaneously by superimposing their signals. Specifically, the transmitted signal from BS$_i$ is represented as $x_i = \sqrt{\zeta_i P_i} s_e + \sqrt{(1 - \zeta_i) P_i} s_{c_i}$, where $P_i$ is the transmit power of BS$_i$, $s_e$ and $s_{c_i}$ are the signals intended for U$^e_f$ and U$^i_c$, respectively, and $\zeta_i$ is the power allocation factor for the edge users. To ensure successful decoding by U$^i_c$, i.e., the strong user, $\zeta_i$ is constrained to $0.5 < \zeta_i < 1$~\cite{obeed2020user}.

The signal received by U$^e_f$ can be expressed as
\begin{multline}
  y^e_f = \underbrace{\sum_{j \in \mathcal{J}} H^e_{j, f} \sqrt{\zeta_j P_j} s_e}_{\text{CoMP gain}} + \underbrace{\sum_{j \in \mathcal{J}} H^e_{j, f} \sqrt{(1 - \zeta_j) P_j} s_{c_j}}_{\text{intra-cluster interference}} \\
  + \underbrace{\sum_{m \in \mathcal{I} \setminus \mathcal{J}} H^e_{m, f} x_i}_{\text{inter-cell interference}} + n_o,
\end{multline}
where $H^e_{j, f} = h^e_{j, f} + \mathbf{h}^T_{\text{R}_j, f} \mathbf{\Theta}_j \mathbf{h}_{j, \text{R}_j}$ and $H^e_{m, f} = h^e_{m, f} + \mathbf{h}^T_{\text{R}_i, f} \mathbf{\Theta}_i \mathbf{h}_{m, \text{R}_i}$ are the effective channels between BS$_j$ and U$^e_f$ and between BS$_i$ and U$^e_f$, respectively, and $n_o \sim \mathcal{CN}(0, \sigma^2)$ denotes the additive white Gaussian noise. To minimize synchronization overhead, we employ non-coherent JT-CoMP, where the edge user U$^e_f$ combines signals from cooperative BSs without CSI exchange~\cite{tanbourgi2014tractable}. Therefore, the signal-to-interference-plus-noise ratio (SINR) at U$^e_f$ is given by
\begin{equation}
  \gamma^e_f = \frac{\sum_{j \in \mathcal{J}} \zeta_j P_j |H^e_{j, f}|^2}{\sum_{j \in \mathcal{J}} (1 - \zeta_j) P_j |H^e_{j, f}|^2 + Y_e + \sigma^2},
\end{equation}
where $Y_f = \sum_{m \in \mathcal{I} \setminus \mathcal{J}} P_m |H^e_{m, f}|^2$ encapsulates the impact of inter-cell interference.

Meanwhile, the signal received by U$^i_c$ is expressed as
\begin{multline}
  y^i_c = h^i_{i, c} \sqrt{(1 - \zeta_i) P_i} s_{c_i} + \sum_{j \in \mathcal{J}} h^i_{j, c} \sqrt{\zeta_j P_j} s_e \\
  + \sum_{j \in \mathcal{J}, j \neq i} h^i_{j, c} \sqrt{(1 - \zeta_j) P_j} s_{c_j} + \sum_{m \in \mathcal{I} \setminus \mathcal{J}} h^i_{m, c} x_m + n_o.
\end{multline}
Based on the SIC principle, the SINR at U$^i_c$ for decoding the signal intended for U$^e_f$ is given by
\begin{equation}
  \gamma^i_{c\rightarrow f} = \frac{\sum_{j \in \mathcal{J}} \zeta_j P_j |h^i_{j, c}|^2}{\sum_{j \in \mathcal{J}} (1 - \zeta_j) P_j |h^i_{j, c}|^2 + Y_i + \sigma^2},
\end{equation}
and the SINR at U$^i_c$ for decoding its own signal is
\begin{equation}
  \gamma^i_c = \frac{(1 - \zeta_i) P_i |h^i_{i, c}|^2}{\sum_{j \in \mathcal{J}, j \neq i} (1 - \zeta_j) P_j |h^i_{j, c}|^2 + Y_i + \sigma^2}.
\end{equation}
It is important to note that due to their placement at the cell edge, the impact of RIS on the channels experienced by U$^i_c$ is negligible. Consequently, the SINR expressions for U$^i_c$ only consider the direct links between the BSs and the users. Finally, the achievable rates for U$^e_f$ and U$^i_c$ can be calculated as
\begin{align}
  \mathcal{R}^e_f & = \log_2(1 + \gamma^e_f), %\,\text{and} \\
\end{align}
and
\begin{align}
  \mathcal{R}^i_{c} & = \log_2(1 + \gamma^i_{c}).
\end{align}

An outage event occurs for cell-center users if U$^i_c$ fails to decode $s_e$ or is capable of decoding $s_e$ but fails to decode $s_{c_i}$. The corresponding outage probability can be defined as $\mathbb{P}^i_c = 1 - \mathbb{P}(\gamma^i_{c\rightarrow f} > \hat{\gamma_f}, \gamma^i_c > \hat{\gamma_c})$, where $\hat{\gamma_f}=2^{\mathcal{R}^e_{th}}-1$ and $\hat{\gamma_c}=2^{\mathcal{R}^i_{th}}-1$ represent the target SINR thresholds for U$^e_f$ and U$^i_c$, respectively, corresponding to their target rates $\mathcal{R}^e_{th}$ and $\mathcal{R}^i_{th}$. For edge user U$^e_f$, an outage occurs if it fails to decode $s_e$, and the outage probability can be formulated as
%$\mathbb{P}^e_f = \mathbb{P}(\gamma^e_f < \hat{\gamma_f})$.
\begin{align}
  \mathbb{P}^e_f = \mathbb{P}(\gamma^e_f < \hat{\gamma_f}).
\end{align}

\subsection{Energy Efficiency}
To evaluate the performance of the network in terms of both spectral efficiency and power consumption, we define energy efficiency as the ratio of the achievable outage sum rate to the total power expended. This metric essentially quantifies whether the gains in the outage sum rate achieved through CoMP transmission outweigh the associated increase in overall power consumption. Mathematically, the energy efficiency is formulated as
\begin{equation}
  \label{eq:eff}
  \eta_\text{EE} = \sum_{i \in \mathcal{I}} \frac{\sum_{c \in \mathcal{C}^{(i)}} \mathcal{R}^i_{\text{out}_c}}{\frac{1}{\lambda}P_i + P_Q} + \sum_{j \in \mathcal{J}} \frac{\sum_{f \in \mathcal{F}} \mathcal{R}^e_{\text{out}_f}}{\frac{1}{\lambda}P_j + P_Q + P_{\text{R}}},
\end{equation}
where $\mathcal{R}^i_{\text{out}_c} = (1 - \mathbb{P}^i_c) \mathcal{R}^i_{c}$ and $\mathcal{R}^e_{\text{out}_f} = (1 - \mathbb{P}^e_f) \mathcal{R}^e_f$ represent the effective achievable rates for U$^i_c$ and U$^e_f$, respectively, taking into account their outage probabilities. Additionally, $P_Q$ represents the static power consumption within a cell, $P_{\text{R}}=KP_\text{ele}$ denotes the cumulative power consumption of R$_i$, where $P_\text{ele}$ is the power consumption of each RIS element, and $\lambda \in (0, 1]$ signifies the power amplifier efficiency.

  \subsection{Passive Beamforming Design}
  The primary objective of the PBF design in this work is to optimize the energy efficiency of the network by strategically adjusting the RIS phase shifts using the EC approach. This involves configuring R$_j$ associated with cooperative BS$_j$ to enhance the signal quality for U$^e_f$, while simultaneously utilizing R$_k$, $\forall m \in \mathcal{I} \setminus \mathcal{J}$, to suppress the inter-cell interference experienced by U$^e_f$. The optimization problem is formally defined as
  \begin{maxi!}|s|
  {\mathbf{\Phi}}{\text{energy efficiency}~\eta_\text{EE}~\text{in}~\eqref{eq:eff}}{\label{eq:opt}}{}
  \addConstraint{\theta^j_k}{\in [-\pi, \pi),}{\; \forall k \in [1, K],\, j \in \mathcal{J}} \label{eq:coop}
  \addConstraint{\theta^m_k}{\in [-\pi, \pi),}{\; \forall k \in [1, K],\, m \in \mathcal{I} \setminus \mathcal{J}} \label{eq:interf}
  \end{maxi!}
  where $\mathbf{\Phi} = \{\mathbf{\Theta}_1, \mathbf{\Theta}_2, \ldots, \mathbf{\Theta}_I\}$ represents the set of phase shift matrices for all RIS.

  For cooperative BSs, the phase shifts of R$_i$ in constraint~\eqref{eq:coop} are optimized to maximize the effective channel gain $|H^e_{j, f}|^2$, thereby improving energy efficiency. From~\cite{wu2019intelligent}, the optimal phase shift for each element of R$_j$ is given by
  \begin{equation}
    \theta^j_k = \arg(h^e_{j, f}) - \arg(h_{\text{R}_j, f}^{(k)} \cdot h_{j, \text{R}_j}^{(k)}),
  \end{equation}
  where $\arg(\cdot)$ denotes the argument function, and $h_{\text{R}_j, f}^{(k)}$ and $h_{j, \text{R}_j}^{(k)}$ represent the $k$-th elements of the channel vectors $\mathbf{h}_{\text{R}_j, f}$ and $\mathbf{h}_{j, \text{R}_j}$, respectively.

  In contrast to signal enhancement, the phase shifts of R$_m$ for non-cooperating BSs in constraint~\eqref{eq:interf} are adjusted to minimize the effective channel gain $|H^e_{m, f}|^2$, thus mitigating interference. Thus, the optimal phase shift for each element of R$_m$ can be calculated as
  \begin{equation}
    \theta^m_k = \text{mod}\left[\phi^m_k + \pi,\: 2\pi\right] - \pi,
  \end{equation}
  where $\phi^m_k = \arg(h^e_{m, f}) - \arg(h_{\text{R}_m, f}^{(k)} \cdot h_{m, \text{R}_m}^{(k)})$ and $\text{mod}[\cdot]$ denotes the modulo operation. Adding $\pi$ ensures a 180° phase shift, effectively suppressing interference.

  It should be noted that incorporating cell-center users into the optimization problem, or increasing the number of users per cell, significantly increases the complexity of the beamforming design. To maintain tractability, this work focuses on a single edge user and a single cell-center user per cell.

  \section{Numerical Results}

  \begin{table}[b]
    \vspace{-1em}
    \centering
    \caption{Simulation Parameters}
    \label{tab:params}
    \begin{tabular}{|c|c|}
      \hline
      \textbf{Parameter}                                    & \textbf{Value} \\
      \hline
      \hline
      Amplifier efficiency $\lambda$                        & $0.4$          \\
      Reference path loss $\rho_o$                          & $-30$ dBm      \\
      Static power consumption $P_Q$                        & $30$ dBm       \\
      Power dissipated at $k$-th RIS element $P_\text{ele}$ & $5$ dBm        \\
      Target cell-center user rate $\mathcal{R}^i_{th}$     & $1$ bps/Hz     \\
      Target edge user rate $\mathcal{R}^e_{th}$            & $0.5$ bps/Hz   \\
      Rician factor $\kappa$                                & $3$ dB         \\
      \# of channel realizations $N_{\text{mc}}$            & $10^4$         \\
      \hline
    \end{tabular}
  \end{table}

  \subsection{Simulation Setup}
  The performance of the proposed design is evaluated in a network consisting of $I = 6$ cells, each with a radius of $R_i = 75$ m. All BSs transmit at the same power level, i.e., $P_i = P_t$ dBm, $\forall i \in \mathcal{I}$, and the power allocation factor for edge users is set to $\zeta_i = 0.7$. The RIS are positioned at the cell edge, resulting in a distance of $d_{i, \text{R}_i} = 75$ m between BS$_i$ and R$_i$. The distances between BSs and users are configured as follows: $d^i_{i, c} = 50$ m, $d^e_{i, f} = 150$ m, and $d^i_{k, c} = 200$ m for $k \neq i$, where $i \in \mathcal{I}$, $c \in \mathcal{C}^{(i)}$, and $f \in \mathcal{F}$. Similarly, the distance between RIS and edge users is set to $d^e_{i, f} = 75$ m.

  The path loss exponents are set to $\alpha_{\text{R}} = 2.7$, $\alpha_i = 3$, $\alpha_e = 3.5$, and $\alpha_{\text{ici}} = 4$ for RIS, BS, edge user, and inter-cell interference links, respectively. The network operates at a carrier frequency of $f_c = 2.4$ GHz, and the noise power is defined as $\sigma^2 = -174 + 10 \log_{10}(B)$, with a bandwidth $B = 10$ MHz. Table~\ref{tab:params} summarizes the remaining simulation parameters.

  \begin{figure}
    \centerline{
      \includegraphics[width=0.925\columnwidth]{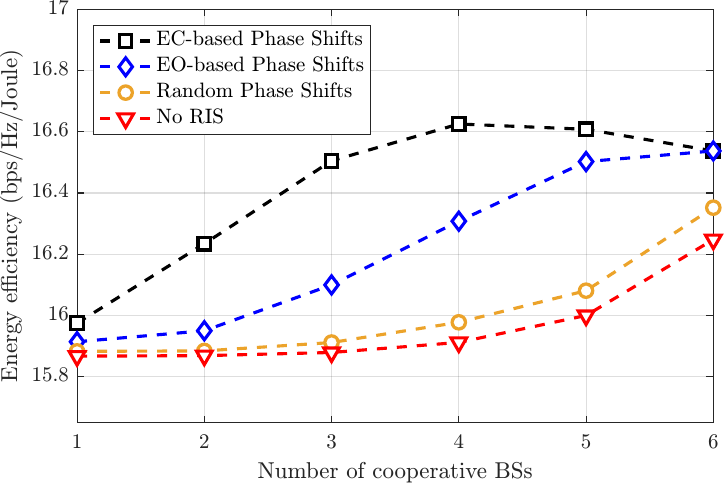}}
    \caption{Energy efficiency vs. number of cooperative BSs $J$ at $P_t = 0$ dBm and under various RIS configurations with $K = 70$ elements.}
    \label{fig:eff}
  \end{figure}

  \begin{figure}
    \centering
    \includegraphics[width=0.925\columnwidth]{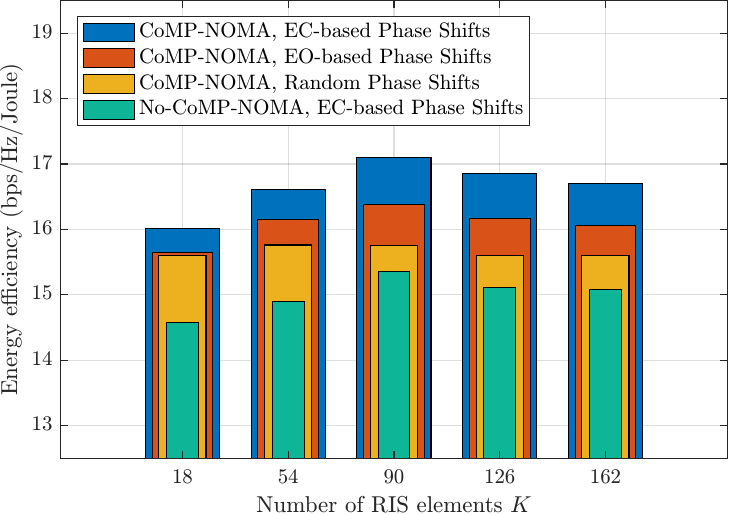}
    \caption{Energy efficiency vs. number of RIS elements $K$ at $P_t = 0$ dBm and $J = 4$ cooperative BSs.}
    \label{fig:bar}
  \end{figure}

  \subsection{Results}
  Fig.~\ref{fig:eff} illustrates the energy efficiency of the network as a function of the number of cooperating BSs $J$ for various RIS configurations. The investigated scenarios include: no RIS, RIS with random phase shifts, EO-based RIS, and EC-based RIS. We observe that for the EO and EC configurations, the energy efficiency initially increases with a growing number of cooperating BSs, reaching a peak at $J = 4$. However, beyond this point, the efficiency experiences a decline. This can be attributed to the saturation of both the achievable rate and the outage probability, leading to diminishing returns with increasing cooperation. It should be noted that the EC configuration consistently outperforms other scenarios $\forall J$, except when $J = I$. In this particular case where all BSs are cooperative, interference cancellation becomes redundant, leading to equivalent performance between the EO and EC configurations.

  The impact of the number of RIS elements $K$ on the energy efficiency is depicted in Fig.~\ref{fig:bar}. A clear trend emerges across all configurations: energy efficiency improves as the number of RIS elements increases, reaching a peak at $K = 90$. Beyond this point, the gains in the outage sum rate are counterbalanced by the increased power consumption associated with additional elements, resulting in a decline in energy efficiency. Notably, the EC configuration consistently outperforms the EO configuration across all values of $K$. Furthermore, the No-CoMP scenario exhibits the lowest energy efficiency for all values of $K$, underscoring the significance of CoMP in enhancing overall network performance.

\begin{figure}
    \centerline{
      \includegraphics[width=0.925\columnwidth]{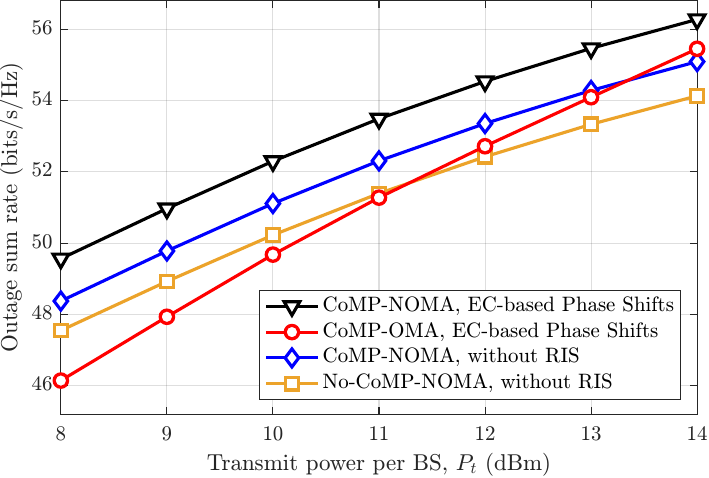}}
    \caption{Outage sum rate vs. transmit power $P_t$ with $J = 4$ cooperative BSs and $K = 70$ elements.}
    \label{fig:osum}
  \end{figure}

  The analysis of the outage sum rate concerning transmit power $P_t$ is illustrated in Fig.~\ref{fig:osum}. To provide a comparative context, orthogonal multiple-access (OMA) is included in the evaluation. As anticipated, the outage sum rate demonstrates an increasing trend with an increase in transmit power across all considered configurations. Notably, the implementation of CoMP-NOMA with EC-based RIS consistently surpasses all other scenarios, including OMA, throughout the entire range of $P_t$ values. An interesting observation is that the CoMP-OMA configuration exhibits steeper increases in outage sum rate per unit of power compared to the CoMP-NOMA configurations. This characteristic can be attributed to the more pronounced rate saturation effect in NOMA networks, wherein the achievable rate is limited by the SIC capabilities of the users.

\begin{figure}
    \centering
    \includegraphics[width=0.925\columnwidth]{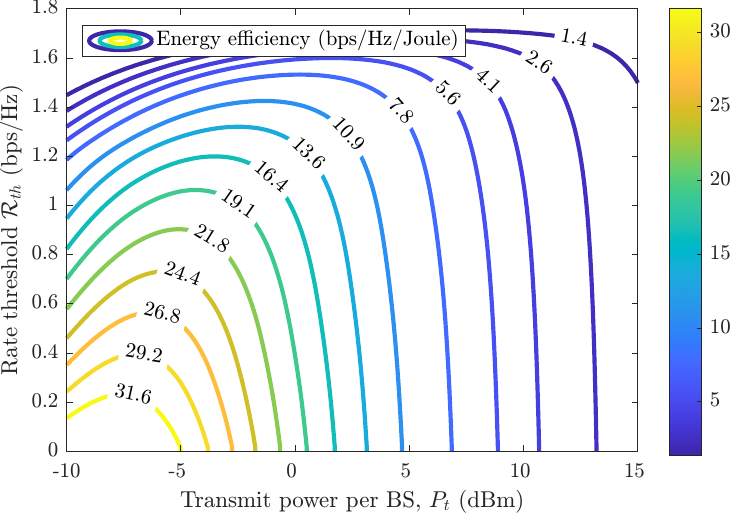}
    \caption{Energy efficiency contour plot for varying transmit power $P_t$ and rate threshold $\mathcal{R}_{th}$ with $J = 4$ cooperative BSs and $K = 70$ elements.}
    \label{fig:cont}
  \end{figure}

  Fig.~\ref{fig:cont} presents a contour plot illustrating the relationship between energy efficiency, transmit power $P_t$, and joint rate threshold $\mathcal{R}_{th}$, where we set $\mathcal{R}^i_{th} = \mathcal{R}^e_{th} = \mathcal{R}_{th}$ for simplicity. Although the highest levels of energy efficiency are found in regions with low $P_t$ and $\mathcal{R}_{th}$, these operating points are often impractical due to minimum user rate requirements. As we move diagonally across the plot, a clear trend emerges: energy efficiency decreases as the rate threshold increases. This inverse relationship highlights the inherent trade-off between achieving higher data rates and maintaining energy efficiency. To meet the demand for higher $\mathcal{R}_{th}$, the network necessitates higher $P_t$, leading to increased power consumption and, subsequently, reduced energy efficiency.

\begin{figure}
  \centering
  \includegraphics[width=0.925\columnwidth]{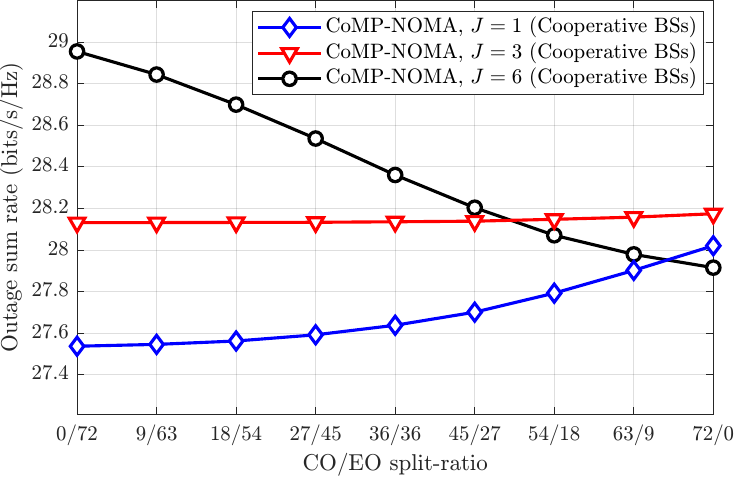}
  \caption{Outage sum rate vs. CO/EO split-ratio for different cooperative BSs $J$ with $P_t = 0$ dBm with $K = 72$ elements.}
  \label{fig:ecco}
\end{figure}

  Lastly, to summarize the trade-off between PBF designs, we analyze the outage sum rate as a function of the ratio of Cancellation-only (CO) and EO elements in Fig.~\ref{fig:ecco}. As expected, increasing CO elements led to a decrease in the outage sum rate for the fully cooperative scenario, $J=I$. However, with half of the BSs cooperating, the outage sum rate remained relatively stable regardless of the CO/EO ratio, demonstrating a balanced contribution from both designs. Interestingly, the highest gains were observed when $J=1$, with all elements employing the CO scheme. This emphasizes the critical role of interference cancellation in such scenarios. Moreover, these findings highlight the importance of optimizing network parameters through a robust optimization framework, a direction we intend to explore in the future.

\section*{Acknowledgement}
%The work of H. Jung was supported by the MSIT, Korea, in part under the National Research Foundation of Korea grants (RS-2023-00303757), in part under the ITRC support programs
%(IITP-2025-RS-2021-II212046), and in part under the Convergence security core talent training business support program (IITP-2023-RS-2023-00266615) supervised by the IITP.
%The work of H. Jung was supported by the Institute of Information \& communications Technology Planning \& evaluation (IITP) grant funded by the Korean government (MSIT) (No. RS-2024-00397480, System Development of Smart Repeater in Upper-mid Band). The work of A. Mahmood and M. Gidlund was supported by the Knowledge Foundation Sweden (KKS) research profile NIIT.
This work of H. Jung was supported by the Institute of Information \& communications Technology Planning \& evaluation (IITP) grant funded by the Korean government (MSIT) (No. RS-2024-00397480, System Development of Smart Repeater in Upper-mid Band). The work of A. Mahmood and M.~Gidlund was supported by the Knowledge Foundation Sweden (KKS) research profile NIIT.

\section{Conclusion}
This paper delved into the investigation of an energy-efficient passive beamforming design of RIS-assisted CoMP-NOMA networks. We introduced two distinct RIS configurations, namely, enhancement-only PBF (EO) and enhancement \& cancellation PBF (EC). Our comprehensive analysis revealed that RIS-assisted CoMP-NOMA networks can significantly improve energy efficiency compared to traditional CoMP-NOMA networks. Additionally, we formulated a PBF design problem aimed at optimizing RIS phase shifts to maximize energy efficiency. The results showcased the influence of various factors on the performance of the network, including the number of cooperating BSs, the number of RIS elements, and the PBF configuration employed.

This work contributes valuable insights for the design and optimization of RIS-assisted CoMP-NOMA networks, paving the way for improved energy efficiency and overall network performance. Future research directions may include exploring more intricate environmental scenarios and investigating the impact of practical RIS limitations, such as phase quantization errors, on the overall system performance. Furthermore, extending the proposed PBF design to accommodate dynamic channel conditions and user mobility presents a promising avenue for further investigation and development.

\bibliographystyle{ieeetr}
% \bibliography{references/ref}

\begin{thebibliography}{10}

  \bibitem{wu2021intelligent}
  Q.~Wu, S.~Zhang, B.~Zheng, C.~You, and R.~Zhang, ``{Intelligent reflecting
  surface-aided wireless communications: A tutorial},'' {\em IEEE Trans.
  Commun.}, vol.~69, no.~5, pp.~3313--3351, 2021.

  \bibitem{zeng2020reconfigurable}
  S.~Zeng, H.~Zhang, B.~Di, Z.~Han, and L.~Song, ``{Reconfigurable intelligent
  surface (RIS) assisted wireless coverage extension: RIS orientation and
  location optimization},'' {\em IEEE Commun. Lett.}, vol.~25, no.~1,
  pp.~269--273, 2020.

  \bibitem{zhang2020capacity}
  Z.~Zhang and L.~Dai, ``{Capacity improvement in wideband reconfigurable
  intelligent surface-aided cell-free network},'' in {\em IEEE 21st Workshop
      Signal Process. Adv. Wirel. Commun. SPAWC}, pp.~1--5, 2020.

  \bibitem{huang2019reconfigurable}
  C.~Huang, A.~Zappone, G.~C. Alexandropoulos, M.~Debbah, and C.~Yuen,
  ``{Reconfigurable intelligent surfaces for energy efficiency in wireless
  communication},'' {\em IEEE Trans. Wirel. Commun.}, vol.~18, no.~8,
  pp.~4157--4170, 2019.

  \bibitem{you2020energy}
  L.~You, J.~Xiong, D.~W.~K. Ng, C.~Yuen, W.~Wang, and X.~Gao, ``{Energy
  efficiency and spectral efficiency tradeoff in RIS-aided multiuser MIMO
  uplink transmission},'' {\em IEEE Trans. Signal Process.}, vol.~69,
  pp.~1407--1421, 2020.

  \bibitem{akyildiz20206g}
  I.~F. Akyildiz, A.~Kak, and S.~Nie, ``{6G and beyond: The future of wireless
  communications systems},'' {\em IEEE Access}, vol.~8, pp.~133995--134030,
  2020.

  \bibitem{liu2017nonorthogonal}
  Y.~Liu, Z.~Qin, M.~Elkashlan, Z.~Ding, A.~Nallanathan, and L.~Hanzo,
  ``{Nonorthogonal multiple access for 5G and beyond},'' {\em Proc. IEEE},
  vol.~105, no.~12, pp.~2347--2381, 2017.

  \bibitem{ding2014performance}
  Z.~Ding, Z.~Yang, P.~Fan, and H.~V. Poor, ``{On the performance of
  non-orthogonal multiple access in 5G systems with randomly deployed users},''
  {\em IEEE Signal Process. Lett.}, vol.~21, no.~12, pp.~1501--1505, 2014.

  \bibitem{yue2018unified}
  X.~Yue, Z.~Qin, Y.~Liu, S.~Kang, and Y.~Chen, ``{A unified framework for
  non-orthogonal multiple access},'' {\em IEEE Trans. Commun.}, vol.~66,
  no.~11, pp.~5346--5359, 2018.

  \bibitem{ali2018downlink}
  M.~S. Ali, E.~Hossain, A.~Al-Dweik, and D.~I. Kim, ``{Downlink power allocation
  for CoMP-NOMA in multi-cell networks},'' {\em IEEE Trans. Commun.}, vol.~66,
  no.~9, pp.~3982--3998, 2018.

  \bibitem{elhattab2020comp}
  M.~Elhattab, M.-A. Arfaoui, and C.~Assi, ``{CoMP transmission in downlink
  NOMA-based heterogeneous cloud radio access networks},'' {\em IEEE Trans.
  Commun.}, vol.~68, no.~12, pp.~7779--7794, 2020.

  \bibitem{elhattab2020reconfigurable}
  M.~Elhattab, M.-A. Arfaoui, C.~Assi, and A.~Ghrayeb, ``{Reconfigurable
  intelligent surface assisted coordinated multipoint in downlink NOMA
  networks},'' {\em IEEE Commun. Lett.}, vol.~25, no.~2, pp.~632--636, 2020.

  \bibitem{hou2021joint}
  T.~Hou, J.~Wang, Y.~Liu, X.~Sun, A.~Li, and B.~Ai, ``{A joint design for
  STAR-RIS enhanced NOMA-CoMP networks: A
  simultaneous-signal-enhancement-and-cancellation-based (SSECB) design},''
  {\em IEEE Trans. Veh. Technol.}, vol.~71, no.~1, pp.~1043--1048, 2021.

  \bibitem{ding2020simple}
  Z.~Ding and H.~V. Poor, ``{A simple design of IRS-NOMA transmission},'' {\em
  IEEE Commun. Lett.}, vol.~24, no.~5, pp.~1119--1123, 2020.

  \bibitem{zheng2022survey}
  B.~Zheng, C.~You, W.~Mei, and R.~Zhang, ``{A survey on channel estimation and
  practical passive beamforming design for intelligent reflecting surface aided
  wireless communications},'' {\em IEEE Commun. Surv. Tutor.}, vol.~24, no.~2,
  pp.~1035--1071, 2022.

  \bibitem{wei2021channel}
  L.~Wei, C.~Huang, G.~C. Alexandropoulos, C.~Yuen, Z.~Zhang, and M.~Debbah,
  ``{Channel estimation for RIS-empowered multi-user MISO wireless
  communications},'' {\em IEEE Transactions on Communications}, vol.~69, no.~6,
  pp.~4144--4157, 2021.

  \bibitem{shtaiwi2021channel}
  E.~Shtaiwi, H.~Zhang, S.~Vishwanath, M.~Youssef, A.~Abdelhadi, and Z.~Han,
  ``{Channel estimation approach for RIS assisted MIMO systems},'' {\em IEEE
  Trans. Cogn. Commun. Netw.}, vol.~7, no.~2, pp.~452--465, 2021.

  \bibitem{zhou2022channel}
  G.~Zhou, C.~Pan, H.~Ren, P.~Popovski, and A.~L. Swindlehurst, ``{Channel
  estimation for RIS-aided multiuser millimeter-wave systems},'' {\em IEEE
  Trans. Signal Process.}, vol.~70, pp.~1478--1492, 2022.

  \bibitem{obeed2020user}
  M.~Obeed, H.~Dahrouj, A.~M. Salhab, S.~A. Zummo, and M.-S. Alouini, ``{User
  pairing, link selection, and power allocation for cooperative NOMA hybrid
  VLC/RF systems},'' {\em IEEE Transactions on Wireless Communications},
  vol.~20, no.~3, pp.~1785--1800, 2020.

  \bibitem{tanbourgi2014tractable}
  R.~Tanbourgi, S.~Singh, J.~G. Andrews, and F.~K. Jondral, ``{A tractable model
  for noncoherent joint-transmission base station cooperation},'' {\em IEEE
  Trans. Wirel. Commun.}, vol.~13, no.~9, pp.~4959--4973, 2014.

  \bibitem{wu2019intelligent}
  Q.~Wu and R.~Zhang, ``{Intelligent reflecting surface enhanced wireless network
  via joint active and passive beamforming},'' {\em IEEE Trans. Wirel.
  Commun.}, vol.~18, no.~11, pp.~5394--5409, 2019.

\end{thebibliography}

\end{document}